\documentclass[journal=tches,preprint]{iacrtrans}

 \usepackage{threeparttable} 
\usepackage{lipsum} 
\usepackage{algorithm}
\usepackage{algorithmic} 
\usepackage{booktabs}

\usepackage{pifont} 

\usepackage{eqparbox}
\renewcommand{\algorithmiccomment}[1]{\hfill\eqparbox{COMMENT}{$\triangleright$ #1}}

\usepackage{amsmath,amssymb}
\usepackage{amsfonts}
\usepackage{amsthm}
\usepackage{bm}
 
\newtheorem*{definition*}{Definition}
\newtheorem*{remark*}{Remark}

\newcommand{\yu}{y_\mathrm{u}}
\authorrunning{Marzougui et al.}
\author{Soundes Marzougui\inst{1} \and Nils Wisiol\inst{1} \and Patrick Gersch\inst{1} \and Juliane Krämer\inst{2} \and Jean-Pierre Seifert \inst{1}}
\institute{
  Technische Universität Berlin, Berlin, Germany \email{{soundes.marzougui, nils.wisiol, patrick.gersch} @tu-berlin.de}
  \and
  Technische Universität Darmstadt, Darmstadt, Germany
  
  \email{juliane@qpc.tu-darmstadt.de}
   
}

\title{Machine-Learning Side-Channel Attacks on the GALACTICS Constant-Time Implementation of BLISS}

\begin{document}

\maketitle

\keywords{side-channel analysis, Machine-Learning, post-quantum cryptography, Gaussian sampler, BLISS, GALACTICS}

\begin{abstract}
 Due to the advancing development of quantum computers, practical attacks on conventional public-key cryptography may become feasible in the next few decades. To address this risk, post-quantum schemes that are secure against quantum attacks are being developed.  
 Lattice-based algorithms are promising replacements for conventional schemes, with BLISS being one of the earliest post-quantum signature schemes in this family. However, required subroutines such as Gaussian sampling have been demonstrated to be a risk for the security of BLISS, since implementing Gaussian sampling both efficient and secure with respect to physical attacks is highly challenging. 
 
 This paper presents three related power side-channel attacks on GALACTICS, the latest constant-time implementation of BLISS. All attacks are based on leakages we identified in the Gaussian sampling and signing algorithm of GALACTICS.
 To run the attack, a profiling phase on a device identical to the device under attack is required to train machine learning classifiers.
 In the attack phase, the leakages of GALACTICS enable the trained classifiers to predict sensitive internal information with high accuracy, paving the road for three different key recovery attacks. 
 We demonstrate the leakages by running GALACTICS on a Cortex-M4 and provide proof-of-concept data and implementation for all our attacks.
\end{abstract}

\section{Introduction}

With the advent of quantum computers, attackers may be able to use Shor's algorithm~\cite{Shorsalgorithm} and break conventional digital signature schemes such as RSA and ECDSA. To address this risk, the development of \emph{post-quantum} cryptographic schemes that are immune to cryptanalytic efforts using quantum algorithms becomes mandatory. 
In the field of post-quantum cryptography, currently, five families of mathematical objects and schemes are discussed: lattice, code, hash, multivariate, and supersingular isogeny-based families. The security of each family relies on its respective underlying cryptographic hardness assumptions. 

The US National Institute of Standards and Technology (NIST) has called for proposals for the standardization of post-quantum cryptographic schemes~\cite{NIST-standardizationround} (i.e., public-key encryption,  digital signatures, and key establishment protocols).
Since five out of seven finalists of the NIST post-quantum cryptography standardization process are lattice-based schemes, lattice-based cryptography can be regarded as the currently most relevant family of post-quantum cryptography. 

BLISS~\cite{latticesignaturebimodal}, one of the earliest lattice-based signature schemes, attracted significant attention from the scientific community. However, despite the emerging real-world adoption~\cite{strongswan} and the efforts targeting efficient and secure implementation~\cite{bartheGALACTICSGaussianSampling2019}, BLISS had been the target of a number of different side-channel attacks~\cite{grootbruinderinkFlushGaussReload2016,tibouchiOneBitAll2020, espitauSideChannelAttacksBLISS2017, toblissbornottobe}.

Security against side-channel attacks is a major concern for schemes that are meant for real-world deployment; the NIST lists the side channel resistance of implementations as one of the criteria for its selection. According to Kocher et al.~\cite{DPAPaulKocher}, side-channel attacks are considered the main threat to cryptographic algorithms meant for deployment in embedded systems. In such attacks, an attacker does not exploit mathematical weaknesses or invalid behavior of an implementation, but uses the behavior of their implementation to reveal secret data.

In the past, BLISS has been attacked by side channel attacks based on cache timing of the CPU, where the attacker exploits the time variation caused by the memory management execution to leak sensitive information \cite{strongswan, bartheGALACTICSGaussianSampling2019, tibouchiOneBitAll2020, toblissbornottobe,espitauSideChannelAttacksBLISS2017}.

Side channel attacks based on power consumption of embedded devices have also gained much research attention. In this class of attacks, the attacker gains information about sensitive internal data of the device by measuring the power a device consumes during a cryptographic operation with, typically with high sample rate. The power side channel attack in this paper can be counted towards the profiling attack category which gained popularity in recent years~\cite{templateattacksinprincipalsubspaces, screamingchannelsmeetradiotransceivers,howdeeplearninghelps,makesomenoise, Maghrebi2016BreakingCI,9217595}. 

\paragraph{Our Contributions} 
We present three machine-learning-based profiling side-channel attacks against the GALACTICS implementation of BLISS, each allowing a full secret key recovery.

\begin{itemize}
    \item We identify four leakages in the power analysis of GALACTICS which allow the prediction of internal values with high accuracy. For one case, we demonstrate the superiority of a prediction based on machine learning over linear regression.
    \item Based on the leakages, we demonstrate three attacks.
    \begin{itemize}
        \item In the first attack, we target the constant-time Gaussian sampler~\cite{zhaoFACCTFAstCompact2020}, and the constant-time sign flip implementation of the signing algorithm. After observing the power consumption of approx. 320 signatures, we are able to fully recover the secret key in only a few seconds.
        \item  In the second attack, we restrict the attacker to only one of the four leakages. Following the strategy due to Groot Bruinderink et al.~\cite{grootbruinderinkFlushGaussReload2016}, we demonstrate that -- at the expense of more observed signatures -- a secret key recovery is still feasible.
        \item Our third attack demonstrates that GALACTICS can also be attacked if the sampling process does not leak any information, i.e., that a secret key recovery by the method of Tibouchi et al.~\cite{tibouchiOneBitAll2020} is still possible even if the attacker only learns information about the sign flipping procedure of the signing algorithm.
    \end{itemize}
    \item Our three attacks are demonstrated via proof-of-concept experiments which were performed on Cortex-M4 micro controllers.
\end{itemize}

\paragraph{Related Work}
The security of lattice-based cryptography relies on the Learning with Errors problem (LWE). An LWE instance contains the secret vectors blinded with a noise vector (error). Usually, the noise vectors are taken from a Gaussian distribution, typically acquiring many samples for a single run of the scheme. The Gaussian sampling process is essential to the security of the scheme, as an attacker with knowledge of the samples can compute the solution to the LWE problem and obtain the secret key of the scheme by solving a system of linear equations. Hence, the Gaussian sampling has been considered a potential weakness of lattice-based schemes in general and BLISS in particular.

The sampling from the Gaussian distribution can be based upon a Cumulative Distribution Table (CDT). The CDT sampler includes a table of cumulative distribution function values that covers a finite interval. To output a sample, one generates a uniform random value and determines the Gaussian sample value by iterating through the CDT until the entry corresponding to the uniform random value is found. A constant-time implementation of a CDT sampler forces the execution of a comparison on all the table's entries~\cite{c28}.

Kim et al. \cite{kimSingleTraceAnalysis2018} found that high-precision Gaussian samples based on CDT are not only inefficient in terms of required storage space, but also showed their insecurity by demonstrating a single trace power analysis attack. By recovering the Gaussian samples, they also break the security of the lattice-based scheme employing the sampler. 

Ducas et al.~\cite{latticesignaturebimodal} proposed a more efficient and secure approach for Gaussian sampling. In a first step, their approach draws from a Gaussian distribution with a small standard deviation. Then, in a second step, these samples are blended with uniformly random values. The deviation of the resulting values from a Gaussian distribution is compensated by an additional rejection condition based on a Bernoulli sample.
Due to the blending with uniform values, even if the CDT sample values are obtained by an attack such as the one by Kim et al.~\cite{kimSingleTraceAnalysis2018}, the values of the final Gaussian sample cannot be derived. Thus, the secret key cannot be recovered as the solution to a system of linear equations. 

Using cache timing attacks, Groot Bruinderink et al.~\cite{grootbruinderinkFlushGaussReload2016} targeted the CDT sampler (a guided table sampler as described in~\cite{guidedtableenhancedlbs}) and the Bernoulli rejection (as described in~\cite{latticesignaturebimodal}) and demonstrated that the obtained leakage along with public information leads to a full recovery of the secret key. 
In a similar side-channel attack, Espitau et al.~\cite{10.1145/3133956.3134028} target the Gaussian sampler and the rejection sampler during the BLISS signing process. Using a branch tracing technique, they reveal the Gaussian samples as well as the Bernoulli samples and demonstrate how to use this information to infer the secret key.  
The techniques demonstrated in these attacks are based on leakages of the Gaussian sampling and the Bernoulli rejection (during Gaussian sampling) and do not apply to BLISS-B, an improved variant of BLISS and the default option in strongSwan~\cite{strongswan}. However, Pessl et al.~\cite{toblissbornottobe} present a
new side-channel key-recovery algorithm against both the original
BLISS and the BLISS-B variant. Their key recovery attack, while based on the same leakages, also works against BLISS-B and recovers the key using, among other tools, integer programs, maximum likelihood tests, and a lattice-basis reduction. 
We conclude that independently of the used techniques for key recovery, the Gaussian sampler and the Bernoulli rejection both pose a risk to the security of BLISS as well as BLISS-B signature schemes.

Gaussian and rejection samplers were not the only vulnerabilities of the BLISS scheme. A recent timing attack against BLISS exploits only the bit sign information to achieve full secret key recovery~\cite{tibouchiOneBitAll2020}. In this attack, Tibouchi and Wallet computed part of the secret key using a maximum likelihood estimation on the space of parameters.
To mitigate this attack, they propose to use a constant-time implementation of sign flip, mitigating the leakage based on cache timing.

To mitigate these side channel attacks, it was suggested to employ a countermeasure consisting of performing a random shuffling (Fisher-Yates random shuffle)~\cite{Saarinen, Knuthrandomshuffle} after using any non-constant-time sampling scheme. The random shuffle is claimed to mask  the  relation between  the  retrieved side-channel information of the samples and the secret ~\cite{Saarinen}. However, this method cannot totally hide the statistical features of the distributions in the attacked vector. An attacker only requires a marginally larger  yet still  practical  number  of  samples  to rearrange the coordinates and reverse the shuffle operation~\cite{peterpesslanalysetheshuffling}.

Another countermeasure is a constant-time sampling for both Gaussian and Bernoulli distribution via value tables of the cumulative distribution function. An example of this approach is the FACCT Gaussian sampler due to Zhao et al.~\cite{zhaoFACCTFAstCompact2020}. Being an extension of the approach suggested by Ducas et al. \cite{latticesignaturebimodal}, the FACCT sampler avoids storing precomputed values for the Bernoulli sampling, using an approach based on polynomial approximation instead.

Taking all known attacks into account, Barthe et al. recently proposed GALACTICS \cite{bartheGALACTICSGaussianSampling2019}, a constant-time implementation of the BLISS signature scheme. The constant-time implementation of GALACTICS' Gaussian sampler employs a similar approach as in~\cite{zhaoFACCTFAstCompact2020}, but extends their approach by avoiding floating-point multiplication, which, on some platforms, are not executed in constant time. To guarantee constant-time evaluation, they suggested using integer polynomials approximation. The GALACTICS implementation presents not only  a constant-time Gaussian sampler but also a constant-time implementation of the whole scheme, including a constant-time implementation of sign flip during signature generation.

\paragraph{Organization} In Sec.~\ref{sec:background}, we introduce the BLISS signature scheme. Subsequently, in Sec.~\ref{sec:overview_of_the_attacks}, we present an overview on the attack and attacker model. In Sec. ~\ref{sec:Experimental_Setup}, we present our experimental setup and explain the different phases of the attacks. In Sec.~\ref{sec:Power_Side_channels_on_GALACTICS}, we analyze four different information leakages of GALACTICS using power analysis. In Sec.~\ref{sec:Secret_key_recovery}, we give a detailed mathematical description of the key recovery strategies. We end the paper with Sec.~\ref{sec:Discussion_and_possible_countermeasures}, where we discuss possible countermeasures against the three proposed attacks.  

\section{Background}
\label{sec:background} 

For any integer $q$, the ring $\mathbb{Z}_{q}$ is represented by the set $[-q/2,q/2) \cap \mathbb{Z}$. Polynomials are defined in the rings $\mathcal{R} = \mathbb{Z}[X]/(X^n+1)$ or $\mathcal{R}_q = \mathbb{Z}_q[X]/(X^n+1)$ and denoted as bold lower case letters.
Vectors are considered column vectors and denoted by bold lower case letters as well, while matrices are denoted by bold upper case letters. By default, we use the Euclidean $L^2$-norm, i.e.
$ \|\bm{v}\|=\sqrt{\sum_{i} v^{2}_i} $.
By $\lfloor x \rceil_d$ we denotes the $d$ highest-order significant bits of an integer $x$, i.e.
$x = \lfloor x\rceil_d \cdot 2^d+x'$, with $x' \in [-2^{d-1},2^{d-1})$.

A \emph{lattice} $\Lambda$ is a discrete subgroup of $\mathbb{R}^{n}$. Given $m\leq n$ linearly independent vectors $\bm{b_1},...,\bm{b_m} \in \mathbb{R}^n$, the lattice $\Lambda (\bm{b_1},...,\bm{b_m})$ is the set of all integer linear combinations of the $\bm{b_i}$'s, i.e.,

\begin{equation}
    \Lambda (\bm{b_1},...,\bm{b_m}) = \Big\{ \sum_{i=1}^{m} x_i\bm{b_i} \mathrel{\Big|} x_i \in \mathbb{Z} \Big\}
\end{equation}

where {$\bm{b_1},...,\bm{b_m}$} is the \emph{basis} of $\Lambda$ and $m$ is the \emph{rank}. In this paper, we consider full-rank lattices, i.e., with $m=n$. An \emph{integer lattice} is a lattice for which the basis vectors are in $\mathbb{Z}^n$. Usually, we consider elements modulo $q$, i.e., the basis vectors and coefficients are taken from $\mathbb{Z}_q$. 

\subsection{Learning with Errors}
The Learning with Errors problem (LWE), which is a generalization of the classical Learning Parities with Noise problem (LPN), was introduced by Oded Regev~\cite{c2}.  

\begin{definition*}  
Let $n$, $q$ be positive integers, and let $\chi$ be a distribution over $\mathbb{Z}$. For $\bm{s} \in \mathbb{Z}^{n}_{q}$, the LWE distribution $A_{\bm{s},\chi}$ is the distribution over $\mathbb{Z}^{n}_{q}$ $\times$ $\mathbb{Z}_{q}$ obtained by choosing $\bm{a}$ $\in$ $\mathbb{Z}^{n}_{q}$ uniformly
at random and an integer error $e$ $\in$ $\mathbb{Z}$ from $\chi$. The distribution outputs the pair $(\bm{a}, \langle \bm{a}, \bm{s}\rangle + e \bmod q) \in \mathbb{Z}^{n}_{q} \times \mathbb{Z}_{q}$.
\end{definition*}  
There are two important computational LWE problems:
\begin{itemize}
    \item The \emph{search problem} is to recover the secret $\bm{s}$ $\in$ $\mathbb{Z}^{n}_{q}$, given a certain number of samples drawn from the LWE \mbox{distribution $A_{\bm{s},\chi}$.}  
    \item The \emph{decision problem} is to distinguish a certain number of samples drawn from the LWE distribution from uniformly random samples. 
\end{itemize} 

For both variants, one often considers two distributions of the secret $\bm{s} \in \mathbb{Z}^{n}_{q}$, first, the uniform distribution, and, second, the distribution $\chi^n \mod q$, where each coordinate is drawn from the error distribution $\chi$ and reduced modulo $q$. The latter is often called the \emph{normal distribution form of LWE}. Since no finite computation can produce a discrete normal distribution, we require sampling  algorithms with bounded running time to sample from distributions that are very close to the desired distribution. The \emph{high}-precision required is governed by the security proof of the crypto scheme. We describe the Discrete Gaussian sampling in the next section.

\subsection{Gaussian Sampling} \label{sec:gaussian-sampling}
For LWE problems, a centered discrete Gaussian distribution is used. It is characterized by two important values: the standard deviation $\sigma$ and the zero mean $\mu$. A value $x \in \mathbb{Z}$ is assigned with probability
$$
\Pr_{X\sim D_\sigma}\left[X = x\right] = \frac{\rho(x)}{  \sum_{y= - \infty}^{\infty} \rho_\sigma(y)},
$$
where $\rho_\sigma(y) = \exp(\frac{-x^2}{2\sigma^2})$
is the continuous Gaussian function.
With $D_{\sigma}^+$ we denote the non-negative part of $D_\sigma$, i.e.,
$$
    \Pr_{X\sim D^+_\sigma}\left[X = x\right] =  \frac{\rho(x)}{  \sum_{y=  0}^{ \infty} \rho_\sigma(y)}.
$$

There are different generic ways to sample from a discrete Gaussian distribution. An early approach employs the cumulative distribution table (CDT) for sampling as described in Alg.~\ref{alg:CDT-sampler} ~\cite{nonuniformrandomvariategeneration}. It consists in computing a table $\Psi$ of cumulative distribution function values of $D_{\sigma}^{+}$ that cover a finite interval [$ - \tau \sigma, + \tau \sigma$ ]. The parameter $\tau$ denotes the tail-cut and is chosen such that the probability for drawing from outside the interval is negligible, e.g., less than $2^{-128}$. To output a sample, one generates a uniformly random value with 128 bits of precision and returns the index of the first entry in the table greater than the random value. 
 
A constant-time implementation of this sampling algorithm requires the execution of a comparison on all the table's entries~\cite{c28}. Moreover, a 128-bit precision table is expensive in terms of memory storage (e.g., 2730 entries for BLISS-I).
 
A more efficient way, proposed in~\cite{c23}, consists of first sampling from a Gaussian distribution $D_{\sigma_{0}}^{+}$ with a small standard deviation $\sigma_0$. Then, those samples are combined to achieve a larger standard deviation $\sigma$. With this approach, the number of CDT entries can be reduced (e.g., to 63 entries for BLISS-I).

An improvement of this approach is used in the constant-time implementation GALACTICS~\cite{bartheGALACTICSGaussianSampling2019}. Here, the Gaussian sampler outputs a sample $y=Kx+y_u$, where $K$ is  a constant, $x$ sampled from $D^+_{\sigma_0}$, and $y_u$ sampled uniformly at random. To obtain samples distributed according to the target discrete distribution $D_{\sigma}$, a rejection condition is applied. This rejection is denoted as the Bernoulli sampling and rejects the sample $y$ with probability $p = \exp(-y_u(y_u+2Kx)/(2\sigma^2))$. To generate negative samples, one can sample and apply a random sign.
This approach decreases the number of required CDT entries further; BLISS-I needs only 10. Hence, this method is memory-efficient and can be implemented in constant time. The cost is dominated by the rejection sampling step and the generation of the uniform randomness.
 
The calculation of the probability $p$ requires high precision and is thus expensive. 
GALACTICS computes the polynomial approximation of the exponential rejection probability $p$ with respect to a relative precision (i.e., the precision is calculated based on the R\'enyi divergence and fixed to 45 bits of relative precision~\cite{bartheGALACTICSGaussianSampling2019}).
 
The idea of polynomial approximation was previously introduced in the description of FACCT sampling algorithm~\cite{zhaoFACCTFAstCompact2020}. The authors of FACCT avoid using floating point division in their polynomial approximation because the division operation is known to not be constant time (depending on the employed platform). They use floating point multiplication instead to compute the exponential probability, but this instruction does not guarantee constant-time execution either~\cite{bartheGALACTICSGaussianSampling2019}. To guarantee constant-time implementation, GALACTICS authors calculate the polynomial approximation using integer polynomials, avoiding floating point multiplication and division altogether.

 \begin{algorithm}
        \caption{Sampling from $D_{\sigma}$}
        \label{alg:gaussian-sample}
        \hspace*{\algorithmicindent} \textbf{Input} Target standard deviation $\sigma$, integer $K$ = $\left\lfloor \frac{\sigma}{\sigma_0}+1 \right\rfloor$, where $\sigma_0= \frac{1}{2 \ln 2}$ \\
        \hspace*{\algorithmicindent} \textbf{Output} A random integer $y$ $\in$ $\mathbb{Z}^+$ according to $D_{K\sigma_0}$ 
        
        \begin{algorithmic}[1]
        \STATE sample $x$ from $D^{+}_{\sigma_0}$
         \algorithmiccomment{Using CDT sampler Alg.~\ref{alg:CDT-sampler}}
        \STATE sample $y_u$ $\in$ $\mathbb{Z}$ uniformly in $\{0,..,K-1\}$
        \STATE sample $b$ with probability $\exp\left(\frac{-y_u(y_u+2Kx)}{2\sigma^2}\right)$
        \STATE if ($b=0$) restart
         \algorithmiccomment{Bernoulli rejection}
        \STATE sample $a$ uniformly in $\{0,1\}$
        \STATE $y \leftarrow (-1)^a \cdot (Kx+y_u)$
         \algorithmiccomment{sign flip}
        \RETURN $y$ 
        \end{algorithmic}
\end{algorithm}

 \begin{algorithm}
        \caption{CDT sampler}
        \label{alg:CDT-sampler}
        \hspace*{\algorithmicindent} \textbf{Input} CDT table $\Psi$ of length $l$, $\sigma$, $\tau$  \\
        \hspace*{\algorithmicindent} \textbf{Output} Random value $x$ according to dist. described by $\Psi$
        \begin{algorithmic}[1]
        \STATE sample uniform random integer $r$ from $[0, \tau \sigma)$
        \STATE $x \leftarrow 0$
        \WHILE{$r > \Psi[x]$}
        \STATE $x \leftarrow x + 1$
        \ENDWHILE
        \RETURN $x$ 
        \end{algorithmic}
\end{algorithm}

\subsection{The BLISS Signature Scheme}
With the current state of the art, BLISS is the most efficient lattice-based signature scheme \cite{inproceedings}. It has been implemented in both software~\cite{latticesignaturebimodal} and hardware~\cite{c18}. BLISS can be seen as a ring-based optimization of the earlier lattice-based scheme of
Lyubashevsky~\cite{c17}, sharing the same ``Fiat–Shamir with aborts'' structure~\cite{c16}.
In a simplified version of the scheme, the public key is an NTRU-like ratio of the form:
$$
        \bm{a_q}  = \bm{s_2} / \bm{s_1} \; \mod \; q,
$$
where the signing polynomials $\bm{s_1}, \bm{s_2} \in \mathcal{R} = \mathbb{Z}[X]/(X^n+1)$ are small and sparse. 
Alg.~\ref{alg:bliss-keygen}  describes the key generation process. The parameters for BLISS are detailed in the specification~\cite{latticesignaturebimodal}.

To sign a message $m \in \{0,1\}^*$, we first generate commitment values $\bm{y_1}, \bm{y_2} \in \mathcal{R}$ with normally distributed coefficients. Then, we compute a hash $\bm{c}$ of the message $m$ together with
$$
    \bm{u}=-\bm{a_qy_1}+\bm{y_2} \mod q
$$
 using a cryptographic hash function modeled as a random oracle taking values in the set of elements of $\mathcal{R}$ with exactly $\kappa$ coefficients equal to 1 and the others to 0. The signature is the triple $(\bm{c}, \bm{z_1}, \bm{z_2})$, where
 \begin{equation}
     \bm{z_i}=\bm{y_i}+\bm{s_ic}.
 \end{equation}
 A rejection condition ensures the independence of $\bm{z_i}$ and $\bm{s_i}$. Verification is possible because $\bm{u}=-\bm{a_qz_1}+\bm{z_2}$.
The full BLISS signature procedure described in Alg.~\ref{alg:bliss-signing} includes several optimizations on top of the above description. In particular, to improve the repetition rate, it targets a Gaussian distribution for the $\bm{z_i}$, so it includes a uniform random choice of their signs. In addition, to
reduce the signature size, the signature element $\bm{z_2}$ is actually transmitted in the compressed form $\bm{z^\dagger}$, and accordingly, the hash input includes only a compressed version of $\bm{u}$. 
For completeness, we also show the verification procedure in Alg.~\ref{alg:bliss-verification}, although we do not use it further in this paper.  
 \begin{algorithm}
        \caption{BLISS key generation}
        \label{alg:bliss-keygen} 
 \hspace*{\algorithmicindent} \textbf{Output} A BLISS key pair ($\bm{A}$,$\bm{S}$) with public key $\bm{A}$= ($\bm{a_1}$,$\bm{a_2}$)$\in \mathcal{R}^2_{2q}$ and secret key \\ $\bm{S}$=( $\bm{s_1}$,$\bm{s_2}$)$\in \mathcal{R}^2_{2q}$ such that $\bm{AS}$= $\bm{a_1}.\bm{s_1}+\bm{a_2}.\bm{s_2}$ $\equiv$
         $q$ mod $2q$ 
        \begin{algorithmic}[1] 
        \STATE choose $\bm{f}, \bm{g} \in \mathcal{R}_{2q}$ uniformly at random with exactly $d_1$ entries in $\{\pm 1\}$ and $d_2$ entries in $\{\pm 2\}$
        \STATE $\bm{S} =(\bm{s_1},\bm{s_2})=(\bm{f},2\bm{g}+1)$
        \STATE if $\bm{S}$ violates certain bounds (details in~\cite{latticesignaturebimodal}), then restart
        \STATE $\bm{a_q} = (2\bm{g}+1)/\bm{f}$ mod $q$ (restart if $\bm{f}$ is not invertible)
        \RETURN $(\bm{A},\bm{S})$ where $A=(2\bm{a_q},q-2)$ mod $2q$
        \end{algorithmic}
\end{algorithm}  
\begin{algorithm}
        \label{algo2}
        \caption{BLISS signature generation}
        \label{alg:bliss-signing}
        \hspace*{\algorithmicindent} \textbf{Input} A message $m$, public key ($\bm{a_1}$,$q-2$), secret key $\bm{S}=(\bm{s_1},\bm{s_2})$ \\
        \hspace*{\algorithmicindent} \textbf{Output} A signature $(\bm{z_1},\bm{z_2^\dagger},\bm{c}) \in \mathbb{Z}^{n}_{2q} \times \mathbb{Z}^{n}_{p} \times \{0,1\}^n$ of the message $m$ 
        
        \begin{algorithmic}[1] 
        \STATE $\bm{y_1},\bm{y_2} \leftarrow D_{\mathbb{Z}^{n},\sigma}$ \algorithmiccomment{Alg.~\ref{alg:gaussian-sample}}
        \STATE $\bm{u}=\zeta. \bm{a_1}.\bm{y_1}+\bm{y_2}$ mod $2q$   
        \algorithmiccomment{$\zeta(q-1) = 1$ mod $2q$}
        \STATE $\bm{c} = H(\lfloor \bm{u} \rceil_d$ mod $p$, $m$)
        \STATE choose a uniform random bit $b$
        \STATE $\bm{z_1}=\bm{y_1}+(-1)^b\bm{s_1}.\bm{c}$ mod $2q$
        \STATE $\bm{z_2}=\bm{y_2}+(-1)^b\bm{s_2}.\bm{c}$ mod $2q$
        \STATE rejection   sampling: \textbf{restart} to step 2 except with probability based on $\sigma$, $\| \bm{Sc}\|$, $\langle \bm{Sc}\rangle $ (details in~\cite{latticesignaturebimodal})
        \STATE $\bm{z_{2}^{\dagger}}=(\lfloor \bm{u} \rceil_d- \lfloor \bm{u}-\bm{z_2}\rceil_d)$ mod $p$
        \RETURN $(\bm{z_1},\bm{z_{2}^{\dagger}},\bm{c})$
        \end{algorithmic}
\end{algorithm}  
\begin{algorithm}
        \caption{BLISS verification algorithm}
        \label{alg:bliss-verification}
        \hspace*{\algorithmicindent} \textbf{Input} Message $m$, public key $\bm{A} = ( \bm{a_1},q-2)$, signature $(\bm{z_1},\bm{z_2^\dagger},\bm{c})$ \\
        \hspace*{\algorithmicindent} \textbf{Output} Accept or reject the signature  
        \begin{algorithmic}[1]  
        \IF{$(\bm{z_1},\bm{z_2^\dagger})$ violates certain bounds  (details in~\cite{latticesignaturebimodal})} 
        \STATE reject
        \ENDIF
        \STATE accept iff $\bm{c} = H( \lfloor \zeta.\bm{a_1}.\bm{z_1}+\zeta.q.\bm{c} \rceil_d+\bm{z_2^\dagger}$ mod $p$, $m)$ 
        \end{algorithmic}
\end{algorithm}
\subsection{GALACTICS Implementation}
An implementation of BLISS with complete timing attack protection -- GALACTICS -- is presented in~\cite{bartheGALACTICSGaussianSampling2019}. The GALACTICS implementation relies on integer arithmetic (limited to addition, multiplication, and shifting on
32-bit and 64-bit operands); division instructions and floating point operations were avoided due to their non-constant-time execution that can present serious security challenges~\cite{c26}.  

GALACTICS authors avoided the previous implementations of sign flips during the signing and sampling processes, such as the original one by Ducas et al.~\cite{latticesignaturebimodal} and the implementation in StrongSwan~\cite{c27}.
These two implementations have conditional branching on the sensitive information about the flipped sign. Therefore, they are not considered constant-time and are vulnerable to timing attacks and branch tracing attacks~\cite{espitauSideChannelAttacksBLISS2017}.

Nevertheless, GALACTICS achieves the same level of efficiency as the original
unprotected code~\cite{latticesignaturebimodal}. It has been proven experimentally, using the \texttt{dudect} by Reparaz et al.~\cite{c29}, that the implementation is constant time and secure against microarchitectural side-channel attacks such as~\cite{grootbruinderinkFlushGaussReload2016, c25}.

\subsection{Machine-learning Model for Profiled Side-Channel Attacks}\label{sec:mlp}
Profiled side-channel attacks are performed in two phases: profiling and attack. The profiling phase can be achieved by creating a template~\cite{templateattacksinprincipalsubspaces, screamingchannelsmeetradiotransceivers}, or training artificial neural networks such as Multilayer perceptron (MLP),  Convolutional Neural Networks, etc~\cite{howdeeplearninghelps,makesomenoise, Maghrebi2016BreakingCI,9217595}.
When using an artificial neural network, the profiling phase consists of training the network to learn the leakage of the target device for all possible values of the sensitive variable. 

In this work, we use leakage models based on MLP classifiers. In each layer of the MLP, the perceptron passes the input into the relu activation function and produces an output.
MLP models consist of at least three layers: the input, hidden, and output layers; multiple hidden layers can be used.
The input layer directly receives the data, whereas the output layer creates the required output. The layers in between are known as hidden layers where the intermediate computation occurs. In the training phase, hidden layers enhance the ability of MLP classifiers to learn a nonlinear function $f:X \rightarrow Y$ by training on data sets $X$ and $Y$. In our setting, $X$ represent the traces captured from the profiling device, while $Y$ are the labels according to the selected leakage model, such as the Hamming weight or value of the desired variable. 
In the attack phase, the trained model (``classifier'') is used to predict the leaked information based on the captured traces from the victim device.

\section{Overview of the Attacks}
\label{sec:overview_of_the_attacks}

Through the analysis of four power analysis leakages in GALACTICS on the Cortex M4, we present three profiling side-channel key recovery attacks. We identify three leakages in the Gaussian sampler implementation, namely in the CDT sampler, the Bernoulli rejection, and the choice of sign of the sample, and one leakage in the signature generation algorithm affecting the sign flip operation.

For the purpose of this study, we consider two identical Cortex M4 CPUs, named Device A and Device B. Device A will be used for profiling, while Device B is the device under attack.

During the profiling phase, we execute the four mentioned functions with random input (according to the respective distribution) on our Device A and collect the corresponding power traces in a controlled environment. We label the power traces with the sensitive internal data that we suspect will be leaked (i.e., whether $\yu = 0$, the value of $x$, and $a$ in Alg.~\ref{alg:gaussian-sample}, and $b$ in Alg.~\ref{alg:bliss-signing}). With the collected data, we train a total of four classifiers, one for each leakage. An overview over the leakages and obtained classifiers is given in Tab.~\ref{tab:leakage}. In a real-world scenario, an attacker would therefore need access to a clone device which has an architecture identical to the device under attack and can be controlled by the attacker to facilitate the collection of training data for the classifieres.

In the attack phase, by observing the power traces of the signature generation on Device B, we use the trained classifiers to predict, with high accuracy, the sensitive internal data of the signing algorithm. Together with known challenge vector elements (i.e., $\bm{z_i}$, $\bm{c}$), we recover the secret key in the first attack as the solution to a system of linear equations, and in the second attack as the kernel of a matrix. 
Additionally, in the third attack, we achieve secret key recovery using the maximum likelihood estimation on a set of signatures.  We detail on the different key recovery methods in Sec.~\ref{sec:Secret_key_recovery}.
An overview of the exploited leakage in the attacks is shown in Tab.~\ref{tab:attack-comparison}.
Hence, to run the attacks, the attacker must be able to trigger the device under attack to generate signatures using a constant secret key. The attacker does not need to choose the signed messages.

\begin{table}[H]
    \centering
    \caption{Comparison of required leaked information (denoted \ding{108}) for different attacks on GALACTICS~\cite{bartheGALACTICSGaussianSampling2019}. The leaked information on $x$ is the value drawn from the discrete Gaussian distribution $D^+_{\sigma_0}$ (with tail cut-off), the leaked information on $a$ and $b$ is the value of the uniform choice from $\{0,1\}$, and the leak on $\yu$ is the information whether $\yu = 0$. For a single signature generation, Alg.~\ref{alg:gaussian-sample} is run at least 512 times; Alg.~\ref{alg:bliss-signing} is run at least once.}
    \begin{tabular}{lccccrr}
    \toprule
    Attack                 & \multicolumn{3}{c}{Alg.~\ref{alg:gaussian-sample}} & Alg.~\ref{alg:bliss-signing} &   required &               \\
                           &        $x$ &        $a$ &      $\yu$ &                                        $b$ & signatures &           \\
    \midrule
    Sec.~\ref{sec:attack1} & \ding{108} & \ding{108} & \ding{108} &                                 \ding{108} &     $\sim$ 320 &   \\
    Sec.~\ref{sec:attack2} &            &             & \ding{108} &                                            &   $\sim$ 2,000 &    \\ 
    Sec.~\ref{sec:attack3} &            &            &            &                                 \ding{108} & $\sim$ 250,000 &   \\ 
    \bottomrule
    \end{tabular}
    \label{tab:attack-comparison}
\end{table}

\section{Experimental Setup}
\label{sec:Experimental_Setup}  

\begin{figure}[h!]
    \centering
    \includegraphics[width=0.7\columnwidth,clip,trim=30 400 100 60 ]{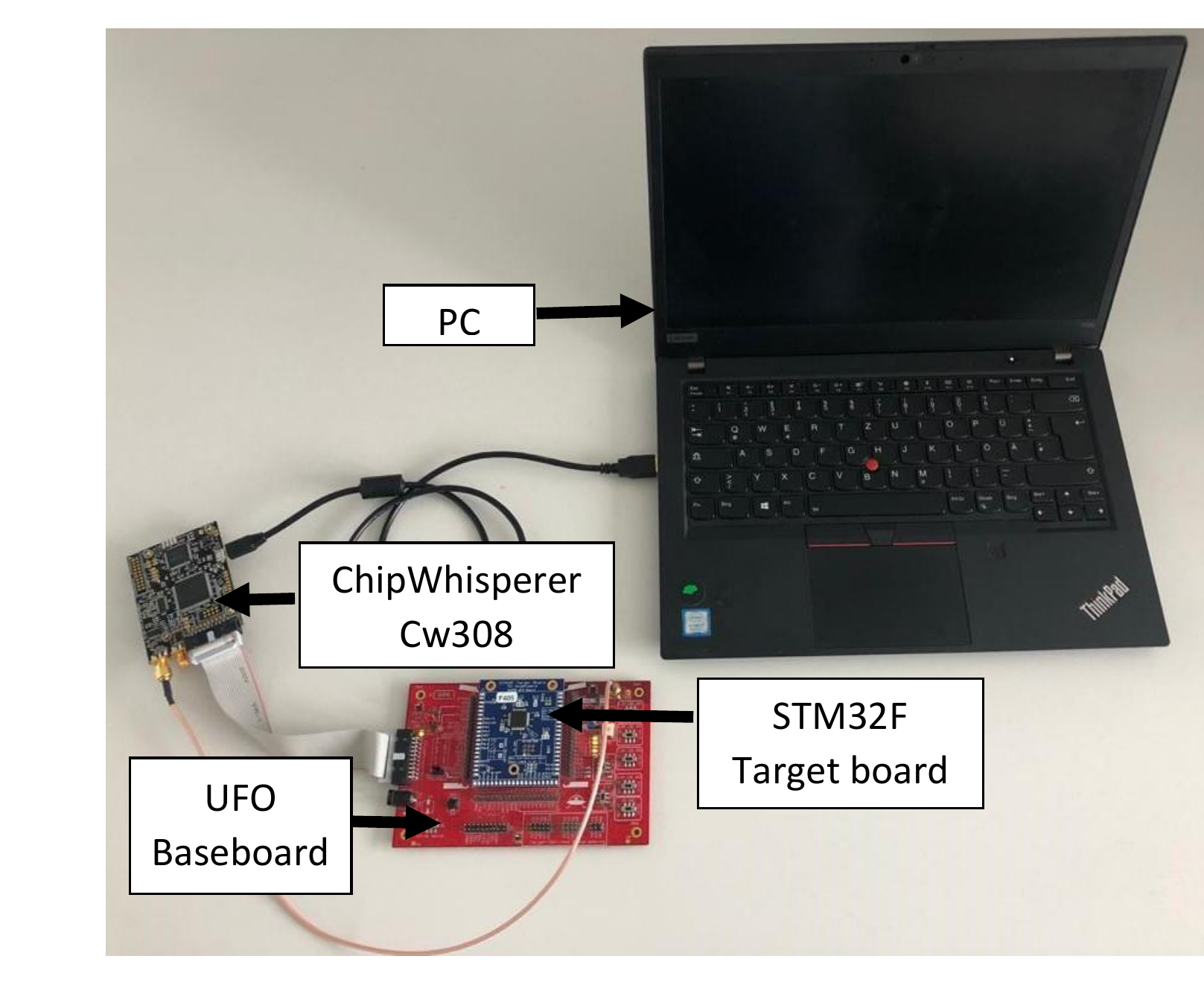}
    \caption{Experimental setup overview}
    \label{fig:workbench}
\end{figure}

\paragraph{Workbench}
To record the traces for profiling and attacking, we used two STM32F4 microcontrollers, Device A and Device B, mounted on a ChipWhisperer Lite CW308 UFO. As the ChipWhisperer Lite is limited to 24,400 samples per recorded trace, it cannot be used to record the entire power trace of the GALACTICS signature generation algorithm. Instead, we traced each of the four targeted leakages individually by first running GALACTICS on an x86 Ubuntu 20.04 server machine. Then, we use an STM32F4 to rerun sections of the GALACTICS code susceptible to leakage and record power traces. During recording, the ChipWhisperer and the microcontroller both run on the same 7,372,800Hz clock. The sampling rate of the ADC was set to 4 samples/cycle with 10-bit resolution and a 45dB low noise gain filter.
Collecting and storing all relevant traces was coordinated using a Python script running on the PC (Fig.~\ref{fig:workbench}). 

\paragraph{Compilation.}
The GALACTICS source code~\cite{c11} was provided as a portable C implementation, which makes it suitable for compilation to different architectures. The original benchmarking was done using SUPERCOP on an Intel Xeon Platinum 8160-based server (Skylake-SP) with Ubuntu 18.04 and gcc 7.3.0.
We compiled GALACTICS using \emph{gcc 8.4.0} on an Intel Core i7-6850K CPU (Broadwell E) running Ubuntu 20.04 and the default SUPERCOP~\cite{SUPERCOP} compiler options (\emph{-march=native -mtune=native -03 -fomit-frame-pointer -fwrapv}). We adapted the flags and added only those for compiling GALACTICS to ARM for usage on the the target M4 device. 

\subsection{Profiling Phase} 
\label{sec:profilingphase}
To prepare the profiling, we signed a number of uniformly random messages using random, individual keys. Then, we collected the internal inputs and outputs of the four functions susceptible to leakage of sensitive data (i.e., CDT sampler, Bernoulli rejection, sign flipping during sampling, and sign flipping during signing), including the randomness used, and stored them along with all public information about the signing process in the profiling database. With this prepared data, we are able to rerun and analyze the parts of the code susceptible to leakage on the profiling device, Device A.

\begin{figure}[hbt!]
    \centering
    \includegraphics[width=0.8\columnwidth,clip,trim=5 230 30 100]{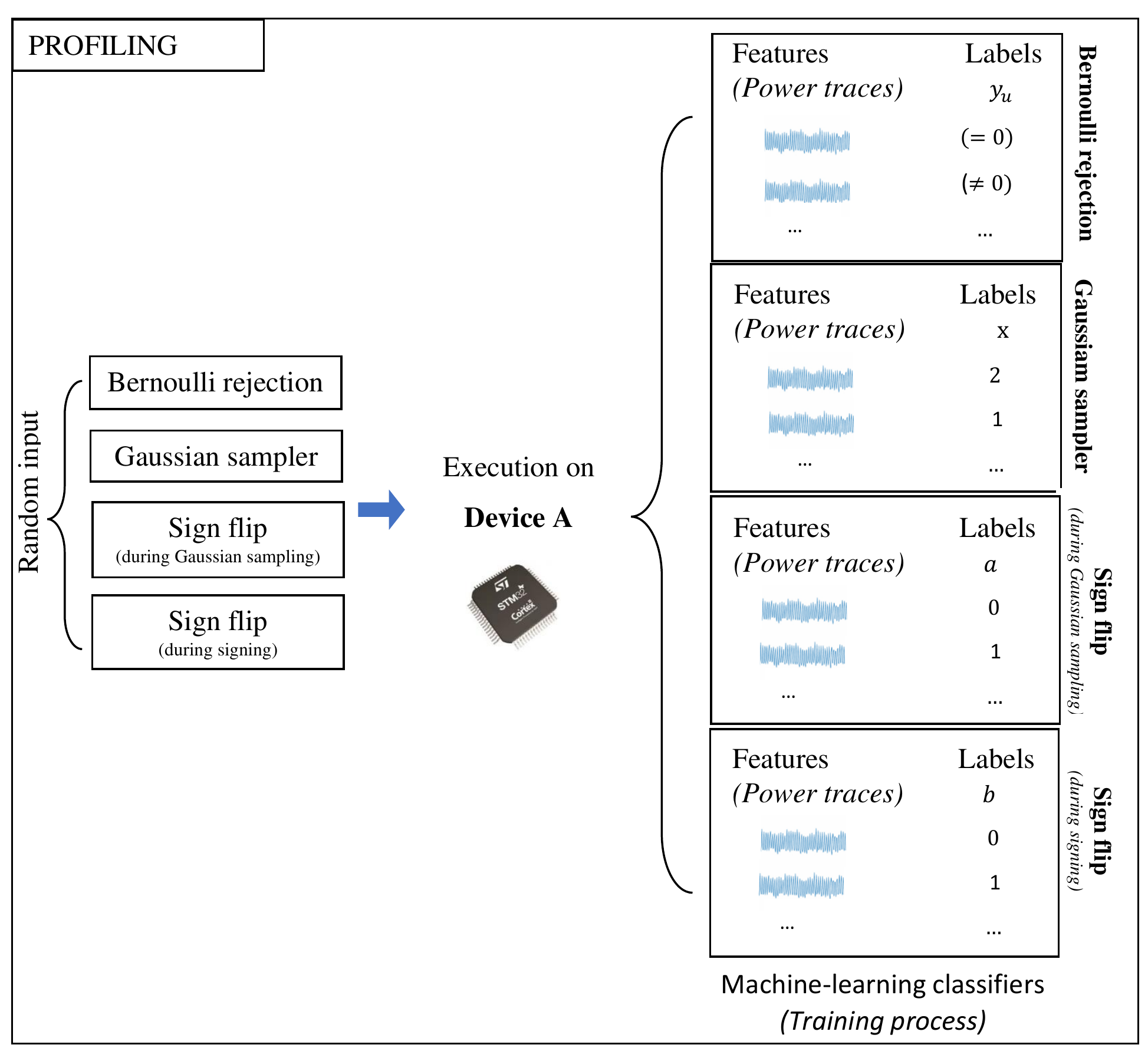}
    \caption{ Profiling phase: Data collection and training of the classifier }
    \label{fig:ml-sc-profiling-step}
\end{figure}  
With the prepared data, we are able to build a list of examples $(x, y) \in \mathbb{R}^t \times Y$, anticipating that the recorded power trace $x$ comprised of $t$ samples leaks information about $y$ from the set of sensitive values $Y$ (the exact values in $Y$ depend on the kind of leakage and are shown in Tab.~\ref{tab:leakage}). The list of these (noisy) examples $(x, y)$ is split into training, validation, and test set of the multilayer perceptron machine-learning model (Sec.~\ref{sec:mlp}), with the power traces $x \in \mathbb{R}^t$ acting as the \emph{features} and the sensitive data $y \in Y$ acting as the \emph{labels}. Note that the noise is limited to the features, while the labels remain noise-free.
To achieve optimal training results, we normalized the feature data to have expectation zero and variance, as it is a common practice in the training of MLP.
The training of the model will result in a function $\hat f: X \to Y$ which approximates $f$. We use the \emph{accuracy} $\Pr_x[f(x) = \hat f(x)]$ as metric for the quality of the approximation. To obtain a valid metric, the accuracy is approximated by evaluating $f$ and $\hat f$ using the test set, whose examples were not used for training.
The choice of model and training parameters can have a decisive influence on the prediction accuracy of the resulting model. We detail on these so-called hyper-parameters in the respective parts of Sec.~\ref{sec:Power_Side_channels_on_GALACTICS}. Fig.~\ref{fig:ml-sc-profiling-step} shows a high-level overview of the profiling phase.
 
\subsection{Attack Phase}
\label{sec:ml-based-sc}
In the attack phase, we generate a number of uniformly random, attacker-known messages. 
Then, we collect the trace snippets of all functions susceptible to leakage of sensitive data (i.e., CDT sampler, Bernoulli rejection, sign flip) and store them along with all public information about the signing process.  
In a real-world attack scenario, the attacker only has access to a complete power trace of the signing process, instead of prepared relevant snippets of it. From a complete trace, however, the position of the relevant snippets can be inferred by a combination of available timing information from GALACTICS (as the implementation is mostly constant time), and correlation of the sampled values with the expectation for the relevant snippet (as the snippets have a distinguished structure).

\begin{figure}[hbt!]
    \centering
    \includegraphics[width=0.8 \columnwidth,clip,trim=5 220 30 100]{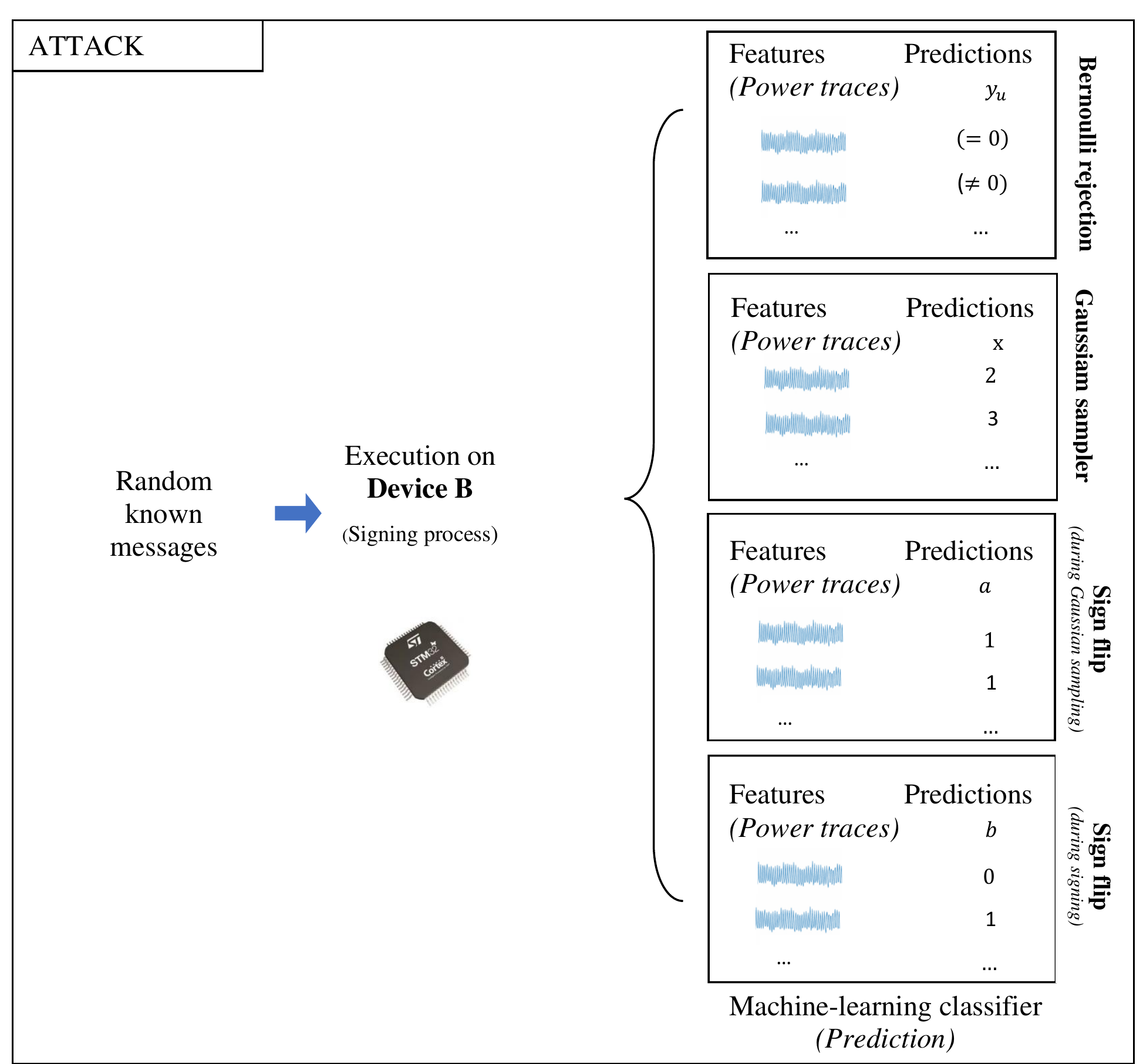}
    \caption{ Attack phase: Tracing the signing on the device under attack and recovering the sensitive data $\yu$, $x$, $a$, and $b$.}
    \label{fig:ml-sc-attack-step}
\end{figure}

With the recorded corresponding power trace snippets $x$ from Device B, the attacker is able to use the trained model $\hat f$ to obtain a prediction $\hat y = \hat f(x)$ of the sensitive data with accuracy as given in Tab. \ref{tab:leakage}. A high-level overview of the attack phase in shown in Fig.~\ref{fig:ml-sc-attack-step}. In Sec.~\ref{sec:Secret_key_recovery}, we explain how to use the leaking information to recover the secret key, in three different attacks.
 
\section{Power Side-Channels Leakage in GALACTICS}
\label{sec:Power_Side_channels_on_GALACTICS}

The code required for training the classifieres described in this section is enclosed with the submission of the paper. It will be published if the paper is accepted. The data required for training has been published at an URL known to the Editors-in-Chief; including the data in the submission is infeasible due to the large size, including the URL in the submission would break anonymity.

\subsection{Adoption of GALACTICS to ARM} 
When studying the leakage of the CDT sampling algorithm (Alg.~\ref{alg:gaussian-sample}, line 1), we noticed that GALACTICS uses short-circuit logic in the table look-up, which leads to a run-time dependency on the sampled value.
These differences do not reveal information to attackers using a timing side channel, as they cannot observe the timing for a single CDT sample, but only for the entire signing process. However, there is little variance of the entire signing process run time caused by the CDT sampling, as each signing process invokes the CDT samples at least 1024 times.
In contrast, using a power side channel allows for a simple power analysis, since the timing information enables an attacker to reconstruct the chosen CDT sample from the power trace with the naked eye.

To accommodate for this, we replaced all short-circuit operations in the CDT sampler with Boolean logic operators, thereby removing all branching instructions. This does not change the functional behavior, while providing a constant run-time of the CDT sampler function. All attacks in this work are against this hardened implementation of the CDT sampler, and we note that the original implementation's leakage of the CDT sampler is considerably easier to recognize.

\begin{table}[]
    \centering
    \begin{tabular}{lllrrr}
        \toprule
        Leakage    &      &            & \multicolumn{3}{c}{accuracy on Device B} \\
        on         & Alg. & Labels $Y$ & trivial & linear regression & MLP  \\
        \midrule
        $x$ & Alg.~\ref{alg:gaussian-sample} & $\{0,1,2,\dots,4\}$ & 77\% & 82.030\% & 93.036\% \\
        $\yu$ & Alg.~\ref{alg:gaussian-sample} & $\{\yu=0,\yu\neq0\}$ & 99.6\% & 99.105\% & ${}^\star$99.953\% \\
        $a$ & Alg.~\ref{alg:gaussian-sample} & $\{a=0,a\neq0\}$ & 50.0\% & 99.801\% & 99.974\%  \\
        $b$ & Alg.~\ref{alg:bliss-signing} & $\{b=0,b\neq0\}$ & 50.0\% &99.671\% & 100\% \\
        \bottomrule
    \end{tabular}  
    \caption{Overview over the GALACTICS power side-channel leakage studied in this work and the performance of our predicted models, trained on data from Device A. Trivial prediction accuracy is achieved by always predicting the most likely label, linear accuracy by using a single-layer perceptron, and MLP accuracy by using a multi-layer perceptron. The MLP classifier for $\yu$ was trained using a custom loss function (see Sec.~\ref{sec:leakage-yu}); the MLP-based classifier for $b$ uses majority vote (see Sec.~\ref{sec:leakage-sign-flip}). }
    \label{tab:leakage}
    \begin{tablenotes}
    \item ${}^\star$ Optimized for a low false positive rate (approx. $10^{-5}$), at the expense of a higher false negative rate, but exceeding linear regression in both false positive (reduction by 38\%) and false negative rate (reduction by 28\%).
    \end{tablenotes}
\end{table}

\subsection{CDT Sampler Leakage on $x$}\label{sec:leakage-cdt-sampler}

\begin{figure}[hbt!]
    \centering
    \includegraphics[width=\columnwidth,clip,trim=0 40 0 5]{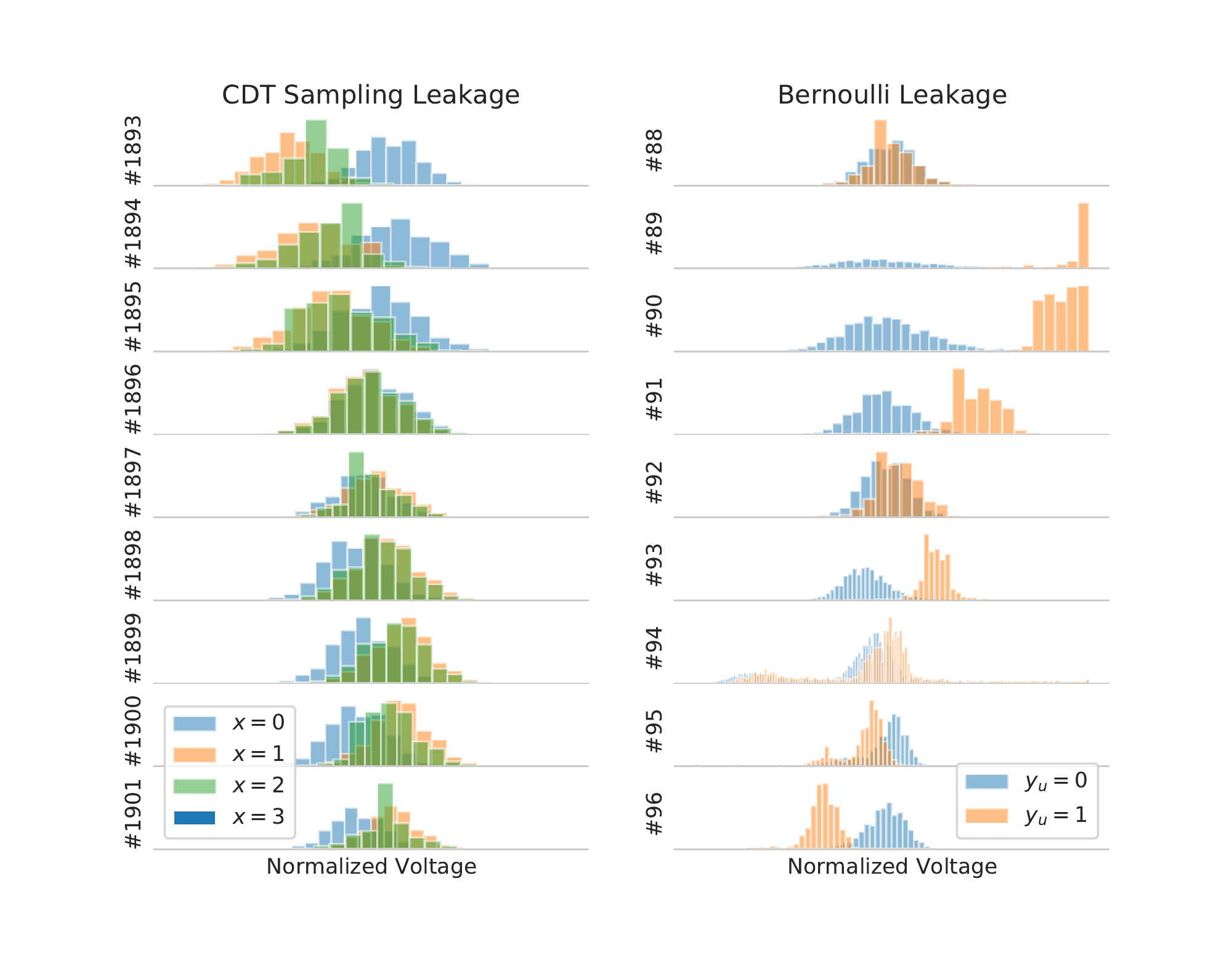}
    \caption{Histograms showing the distributions of the normalized measured voltage during the execution of the GALACTICS CDT Sampler (Alg.~\ref{alg:gaussian-sample}, line 1) and Bernoulli rejection sampling, colored by returned sample.}
    \label{fig:leakage-hist}
\end{figure}

The first leakage of the GALACTICS signing procedure studied in this work is the foundation of the Gaussian sampling process: the sampling from the cumulative distribution table (CDT), as invoked in Alg.~\ref{alg:gaussian-sample}, line 1.
The invoked CDT sampler chooses a value $x \in\mathbb{Z}$ according to the distribution $D^+_{\sigma_0}$, that is, having quickly decreasing probability for larger $x$, with $x=0$ occurring with probability approx. 77\%, and $x=1$ with probability approx. $22\%$, and $x=3$ with probability approx. $2\%$.
The value is determined by a table look-up based on 128 bits of uniform randomness and then shifted 8 bits to facilitate the multiplication with $K = 256$.
In our setup, the tracing of the GALACTICS CDT sampler \href{https://github.com/espitau/GALACTICS/blob/79141a6b4032dbeb79fa8cab0d3c715c0c3eb442/Artifacts/CT-Implementation/bliss1\_galactics/ref/gaussian_sampling.c\#L10}{\texttt{sample\_berncdt()}} including the shift involved 383 cycles, i.e., 1532 measurements using the oscilloscope. The measurements relevant for the recovery of $x$ are shown in Fig.~\ref{fig:leakage-hist}.

With the training set built as outlined in Sec.~\ref{sec:ml-based-sc}, an MLP classifier can be trained to predict the CDT samples based on the leakage. We observed that even with no hidden layers in the MLP, i.e., using plain linear regression, the samples can be predicted correctly in 82\% of cases, a significant increase over the trivial guessing probability of 77\%. By the introduction of a hidden layer, i.e., by also considering nonlinear relationships of power usage and sample value, the accuracy can be increased to 93\%. This observation demonstrates that in our case, the usage of MLP is superior to more traditional power analysis methods. Still, even when using a multilayered model, values $x>3$ are never predicted. Due to the low frequency of these values in the training data, it is difficult to train the model for correct prediction of these values. Nevertheless, even if predicted correctly, the overall accuracy improvement will be negligible.

\subsection{Bernoulli Rejection Leakage on $\yu$}\label{sec:leakage-yu}

After the value $\yu$ is sampled uniformly random from $\{0,\dots,255\}$, in Alg.~\ref{alg:gaussian-sample}, line 3, the \href{https://github.com/espitau/GALACTICS/blob/79141a6b4032dbeb79fa8cab0d3c715c0c3eb442/Artifacts/CT-Implementation/bliss1\_galactics/ref/gaussian_sampling.c\#L131-L133}{Bernoulli rejection condition} is applied, which requires the computation of the value $p$,
$$
   p = \exp(-\yu(\yu+2Kx)/(2\sigma^2)).
$$
While we failed to predict the exact value of $\yu$ from the leakage of this computation, our trained classifier can reveal whether $\yu=0$. We suspect that this is caused by the distinguished Hamming weight of the zero value.  
 
As $\yu=0$ only occurs in $1/256 \approx 0.4\%$ of cases, the training of a highly accurate classifier requires some fine-tuning: a classifier always predicting $\yu\neq0$ already achieves $99.6\%$ accuracy. To prepare the neural network for this task, we adjusted the prediction bias of the output layer to the expected value, $0.4\%$, and weighed the training example according to their label frequency, i.e., we gave examples for $\yu=0$ higher significance in the training process than examples for $\yu\neq0$.

In contrast, we were unable to build a linear regression classifier using a standard binary cross-entropy loss function achieving better accuracy than the trivial $99.6\%$. Our attempts either suffered from predicting many false positives, i.e., predicting $\yu=0$ often when actually $\yu\neq0$, or from just always predicting $\yu\neq0$.

This situation can be alleviated by using an MLP-based classifier with three hidden layers, i.e., also considering nonlinear relationship of leakage and $\yu=0$ \emph{and} tuning the loss function to heavily penalize false positives\footnote{We found that tuning the loss function, but using no hidden layers would suffice for Attack 1, but not for Attack 2. Just using hidden layers, but standard loss, cannot provide predictions accurate enough for Attack 1 or Attack 2.}. To facilitate the latter, we adjusted the training algorithm to compute the binary cross-entropy losses for examples falsely classified as $\yu=0$ or $\yu\neq0$ and for examples correctly classified as $\yu=0$ and $\yu\neq0$ separately. We defined the total loss to be the weighed sum of the four individual losses, with the false positive loss weighed with factor 1000 and the other three with factor 1. Using this technique, we can train a classifier with an extremely low false-positive rate, which is particularly important for our attack in Sec.~\ref{sec:attack1}. Of the 1,450,492 traces collected from Device B, only 15 were falsely predicted to have $\yu=0$. 

We display the measurements responsible for the leakage in Fig.~\ref{fig:leakage-hist}.

\subsection{Sign Flip Leakage on $a$ and $b$}\label{sec:leakage-sign-flip}\label{scb}

The third and fourth leakages studied in this work occur when the sign of an integer is flipped. GALACTICS features a \href{https://github.com/espitau/GALACTICS/blob/79141a6b4032dbeb79fa8cab0d3c715c0c3eb442/Artifacts/CT-Implementation/bliss1\_galactics/ref/bliss.h\#L9}{constant-time implementation of this operation} that mitigates earlier timing and cache side-channel attacks~\cite{tibouchiOneBitAll2020}. Given any integer $x$ and a bit $\lambda \in \{0,1\}$ indicating if the sign of $x$ will be flipped, the value of $(-1)^\lambda\cdot x$ is computed using $${( x \wedge \lambda ) \vee ( (-x) \wedge (1-\lambda) )}$$ to avoid any branching instructions. $\vee$ and $\wedge$ refer to bitwise ``OR'' and ``AND'' operations, respectively.

Again, the power usage profile of the instructions that compute the above expression are leaking information on the resulting sign of $x$. This affects the sign flips in both Alg.~\ref{alg:gaussian-sample}, line 6 and in Alg.~\ref{alg:bliss-signing}, line 5. We note that the inputs to the two sign flip operations have different properties. During the Gaussian sampling (Alg.~\ref{alg:gaussian-sample}, line 6) the input is always positive. Therefore, a flipped sign yields negative results. Moreover, the sign flip operation is called 512 times per signature generation, each using a fresh uniformly random bit indicating the sign. On the other hand, the input to the sign flip operation during the signing process (Alg.~\ref{alg:bliss-signing}, line 5.) can operate on a positive or negative value for $x$, but is called 512 times using the same sign indicator bit. To account for both situations, we use two separate classifiers.

In both cases, using linear regression, the leakage of the sign flip procedure can be used to predict the sign of the result with accuracy approx. 99.7\%, with the prediction of $a$ being slightly more accurate. Similar to the leakage of the CDT sampler in Sec.~\ref{sec:leakage-cdt-sampler}, the accuracy can be further improved to about $99.9\%$ by using an additional hidden layer in the classifier.

As the value of $b$ in Alg.~\ref{alg:bliss-signing}, line 5, is used 512 times, the prediction accuracy of $b$ can be further improved by conducting a majority vote on the classifications of the 512 individual leakages, resulting in a practically perfect prediction for $b$.

\section{Secret Key Recovery}
\label{sec:Secret_key_recovery} 
We present the mathematical foundations of three side channel attacks on BLISS. Each attack is founded on a secret key recovery algorithm that takes as input one or more of the leakages described in Sec.~\ref{sec:Power_Side_channels_on_GALACTICS} and outputs the secret key.  

The first attack, presented in Sec.~\ref{sec:attack1}, targets the entire signing process, i.e., uses all leakages described in Sec.~\ref{sec:Power_Side_channels_on_GALACTICS}. The attacker is hence able to predict, with certain accuracy, the values of the CDT samples $x$, the indicators $a$ for the signs of $y$, whether $\yu=0$ in Alg.~\ref{alg:gaussian-sample}, and the indicators $b$ for the sign flips in Alg.~\ref{alg:bliss-signing}, line 5. By employing the predicted values, we demonstrate how to build a system of linear equations, such that its solution is the secret key. The leakage of approx. 320 calls to the signature generation algorithm is sufficient to obtain a solution.

In the second attack, Sec.~\ref{sec:attack2}, we assume that fewer leakages exist and only information about Bernoulli rejection during the Gaussian sampling to obtain information on whether $\yu=0$ as described in Sec.~\ref{sec:leakage-yu} is available to the attacker. We acknowledge the work of Groot Bruinderink et al. ~\cite{grootbruinderinkFlushGaussReload2016} and extend it to work on the basis of the machine-learing classifier we trained (Sec.~\ref{sec:leakage-yu}) and apply it to the hardened constant-time implementation of BLISS. With 2000 calls to the signature generation algorithm using the key under attack, we construct a matrix whose kernel space is the secret key.

The third attack, presented in Sec.~\ref{sec:attack3}, consists of recovering the sign flip indicators $b$ of Alg.~\ref{alg:bliss-signing} during the  signing process. Then, by following the approach of Tibouchi and Wallet~\cite{tibouchiOneBitAll2020}, the secret key recovery is carried out using a maximum likelihood estimation. The approximate needed number of signatures is 250,000.

We note that to break the BLISS scheme, it is sufficient to extract $ \bm{s_{1}}$, the first half of the secret key used in the signing process, as $\bm{s_2}$ can be recovered through the linear relation:
$$
   \bm{As} = \bm{a_1.s_1} + \bm{a_2.s_2} \equiv q\mod2q.
$$
 Hence, throughout this section, we use $\bm{s} = \bm{s_1}$ and $\bm{z} = \bm{z_1}$. 
 We utilize the indices $i$, $k$ to refer to the $i$-th coefficient and $k$-th signature, respectively, with $i<n$ and $n=512$ for BLISS-I. 


\subsection{Attack 1: CDT Samples, Partial Information on Uniform Samples, and Sign Flips} \label{sec:attack1}   

This attack uses all leakages presented in Sec.~\ref{sec:Power_Side_channels_on_GALACTICS} and thereby requires relatively few executions of the signing algorithm on the device under attack.
After collecting the power traces of approx. 320 runs of the signature generation algorithm, the attacker identifies the relevant snippets. Then, in the first step, an attacker uses the trained classifier of Sec.~\ref{sec:Power_Side_channels_on_GALACTICS} to predict for each call to the Gaussian sampling algorithm Alg.~\ref{alg:gaussian-sample} whether $\yu = 0$, using the power trace snippets of the Bernoulli rejection. As $\yu$ was chosen uniformly at random, we have $\Pr[\yu=0] \approx 0.4\%$. The prediction of $\yu=0$ was optimized to yield few false positives, cf. Sec.~\ref{sec:leakage-yu}.

In this attack, we are interested in runs of the Gaussian sampling algorithm (Alg.~\ref{alg:gaussian-sample}), as in case $\yu=0$, the signature generated in Alg.~\ref{alg:bliss-signing}, line 5, can be written as:
\begin{equation}
    z_{k,i}=(-1)^{a_{k,i}}\cdot Kx_{k,i}+(-1)^{b_{k}} \cdot \langle \bm{s, c_k}\rangle \label{eqn:signature-yu-zero}
\end{equation}
where $a_{k,i}$ is the corresponding sign flip indicator of the Gaussian samples (Alg.~\ref{alg:gaussian-sample}, line 6).
In a second step, the attacker uses the classifiers of Sec.~\ref{sec:leakage-cdt-sampler} and Sec.~\ref{sec:leakage-sign-flip} to obtain high-accuracy predictions for the values $a_{k,i}, b_k$, and $x_{k,i}$ and rearranges Eqn.~\ref{eqn:signature-yu-zero} to:
\begin{equation}
    \langle \bm{s}, \bm{c_k} \rangle = (-1)^{b_k} \left( z_{k,i} - (-1)^{a_{k,i}} Kx_{k,i} \right) \label{eqn:signature-yu-zero-rearranged}
\end{equation}
We note that in this representation, the attacker is able to compute a prediction of the right-hand side of Eqn.~\ref{eqn:signature-yu-zero-rearranged}. With the prediction of the right-hand side and the knowledge of the public value $\bm{c_k}$ the attacker repeats the process to obtain Eqn.~\ref{eqn:signature-yu-zero-rearranged} for different runs of the singing algorithm $k$ and different coefficients $i$. Then, the attacker is able to build a system of linear equations with the coefficients of the secret key $\bm{s}$ acting as unknowns. To that end, the attacker arranges the values of $\bm{c_k}$ as rows in a matrix $\bm{M}$ and the predicted right-hand side values from Eqn.~\ref{eqn:signature-yu-zero-rearranged} as entries in a vector $\bm{r}$ to obtain
\begin{equation}
     \bm{M} \cdot \bm{s} =\bm{r} \label{eqn:attack1-sle}
\end{equation}
A formal algorithmic description of the attack procedure is given in Alg.~\ref{alg:attack1}.

\begin{algorithm}[hbt!]
    \caption{GALACTICS Secret Key Recovery Attack 1}
        \hspace*{\algorithmicindent} \textbf{Input} Relevant power trace snippets of approx. 320 signatures ($T_{y_u},T_{x}, T_{c}, T_{b}$) of linearly independent messages executed on the victim's machine with the same secret key $\bm{s}$; trained classifiers from Sec.~\ref{sec:Power_Side_channels_on_GALACTICS}; input parameters $n$, $\sigma$, $q$, $\kappa$ of BLISS-I (as in~\cite{bartheGALACTICSGaussianSampling2019}).\\
    \hspace*{\algorithmicindent} \textbf{Output} Secret key $\bm{s}$
    \begin{algorithmic}[1] 
    \label{attack1}
    \label{alg:attack1} 
    \STATE $\bm{M} \leftarrow []$,  $\bm{r}\leftarrow[]$, $k\leftarrow 0$
    \FOR{all available traces, indexed by $k$}
    \FOR{each $i = 0..n$} 
    \STATE $\hat y_{u_{k,i}}$ $\leftarrow$ $\yu$-classifier($T_{y_{u_{k,i}}}$)
    \IF{$\hat y_{u_{k,i}}$ = 0}
    \STATE $\hat x_{k,i}$ $ \leftarrow$  $x_{k,i}$-classifier($T_{x_{k,i}}$)
    \STATE $\hat c_{k,i}$ $ \leftarrow$  $c$-classifier($T_{c_{k,i}}$)
    \STATE $\hat b_{k}$ $\leftarrow$  $b$-classifier($T_{b_{k}}$) 
    \STATE add $\bm{c_k}$ as a row to the matrix $\bm{M}$
    \STATE add $(-1)^{\hat b_k} \left( z_{k,i} - (-1)^{\hat a_{k,i}} {K \hat x}_{k,i} \right)$ to $\bm{r}$
    \ENDIF
    \ENDFOR
    \ENDFOR
    \STATE  the secret key $\bm{s}$ satisfies $\bm{M}\cdot\bm{s} = \bm{r}$ approximately, depending on classifier accuracy
    \end{algorithmic}
\end{algorithm} 

Even though the classifiers presented in Sec.~\ref{sec:Power_Side_channels_on_GALACTICS} provide high accuracy, it is unlikely that the system in Eqn.~\ref{eqn:attack1-sle} has an exact solution. As the number of $y_{u_{k,i}}=0$ in the recorded traces cannot be predicted accurately, the system in Eqn.~\ref{eqn:attack1-sle} is likely to be over-determined or under-determined.

To accommodate for this, we use a method that employs the Penrose inverse of $\bm{M}$ to obtain a least-squares solution to the system.  
Using the least-squares method, we found that 499 equations are sufficient for full key recovery if no $y_{u_{k,i}}=0$ false positive is involved.  The reason for the importance of the number of $y_{u_{k,i}}=0$ false positives lies in the nature of Eqn.~\ref{eqn:signature-yu-zero}, which only applies if $y_{u_{k,i}}=0$. The involvement of an equation derived from values when actually $y_{u_{k,i}}\neq0$ has thus dramatic influence on the correctness of the approximate solution to the system. 
 
The noise introduced by the other classifiers seems less important to the overall stability of the approximate solution.

We also note, as  $\yu=0$ occurs with probability $1/256$ in each of the 512 trials per run of the signing algorithm, with perfect prediction of the $\yu=0$ condition, 277 runs of the signing algorithm are sufficient to obtain the 499 required equations. Our attack needs significantly more than that as the classifier for $\yu=0$ prediction was trained for a low rate of false positives, resulting in a higher rate of false negatives.
 
\subsection{Attack 2: Partial Information on Uniform Samples}\label{secondattack}\label{sec:attack2}

This attack is an adaptation of the cache side channel attack by Groot Bruinderink et al.~\cite{grootbruinderinkFlushGaussReload2016}. The only leakage used is the information whether $\yu=0$.
For different signatures $k$, when $ y_{u_{k,i}}  = 0$, the possible values of $y_{k,i}$ (Alg.~\ref{alg:bliss-signing}, line 5) are $\{0, \pm K, \pm 2K, ...\}$. If additionally, $z_{k,i} \in \{0, \pm K, \pm 2K, ...\}$, we can conclude that $\langle  \bm{s}, \bm{c_k} \rangle = 0$. In this case, similarly to the previous attack, we construct a matrix $\bm{M}$ by adding the $\bm{c_k}$ vector as a new row.

As the signing of messages usually operates on a cryptographic hash of the actual message to be signed, we can assume that for different messages, the $\bm{c_k}$ are uniformly random. Hence, the probability that all collected $\bm{c_k}$ are linearly independent is very high. In this case, the kernel space of $\bm{M}$ contains precisely the secret key $\bm{s}$ if $\bm{M}$ has full rank 511. We found that when using a noise-free side channel, 1673 signatures are sufficient to recover the key, but we remark that when fewer signatures are available, this can be compensated by a brute-force search in the (then larger) kernel space of $M$.

To accommodate for potential noisy entries in the matrix $\bm{M}$ caused by false positive predictions of $y_{u_{k,i}}=0$, a larger set of equations can be built. From this larger set, a uniformly random subset of 511 equations can be selected to compute the kernel space. This process can be repeated until 511 noise-free equations were selected, in which case the secret key will be revealed. Using 2000 signatures and the classifier from Sec.~\ref{sec:Power_Side_channels_on_GALACTICS}, we obtained the full secret key after selecting the second random row subset of $\bm{M}$; the total run time of the attack (excluding the collection of traces) was below one minute.
 
\begin{algorithm}[hbt!]
    \caption{GALACTICS Secret Key Recovery Attack 2}
    \hspace*{\algorithmicindent} \textbf{Input} Relevant power trace snippets of 2000 signatures ($T_{y_u}$) of different messages executed on the victim's machine with the same secret key $\bm{s}$; trained classifier from Sec.~\ref{sec:ml-based-sc}; input parameters $n$, $\sigma$, $q$, $\kappa$ of BLISS-I (as in~\cite{bartheGALACTICSGaussianSampling2019}).\\
    \hspace*{\algorithmicindent} \textbf{Output} Secret key $\bm{s}$
    \begin{algorithmic}[1] 
    \label{attack2} 
    \STATE $\bm{M} \leftarrow []$
    \FOR{all available traces, indexed by $k$}
        \FOR{each $i \in \{0,\dots,n\}$} 
            \STATE $\hat{y}_{u_{k,i}}$ $\leftarrow$ classifier($T_{y_{u_{k,i}}}$)
            \IF{$\hat{y}_{u_{k,i}} = 0 \textbf{ and } z_{k,i} \in \{0, \pm K, \pm 2K,..\}$}  
                \STATE add $\bm{c_k}$ as a row to the matrix $M$
            \ENDIF
        \ENDFOR
    \ENDFOR
    \WHILE{$\bm{s}$ not yet found}
        \STATE $\bm{M}' \leftarrow$ random selection of 511 rows of $\bm{M}$
        \STATE check if kernel space of $\bm{M}'$ contains the secret key $\bm{s}$
    \ENDWHILE
    \end{algorithmic}
\end{algorithm} 
 
\subsection{Attack 3: Sign Flip during Signature Generation }\label{sec:attack3}
In this section, we present a third attack that can be mounted even if the sampling process does not leak any data. Following Tibouchi and Wallet \cite{tibouchiOneBitAll2020}, our attack is based solely on the knowledge of $b_k$ and allows the secret key recovery with approx. 250,000 signatures.
In each signature, the sign flip indicator $b_k$ is sampled once and used 512 times; the classifier outlined in Sec.~\ref{sec:leakage-sign-flip} can predict the value of $b_k$ with high accuracy.

To obtain the secret key, the attack formulates the log-likelihood of secret keys based on the recovered values of $b_k$ for a large number of signatures. This function admits a unique maximum at the correct secret key $\bm{s_1}$ and can be reliably obtained using gradient ascent.

\newif\iflong
\longfalse

\iflong
In the following, we introduce the log-likelihood function and prove that it admits a maximum which is the secret key $\bm{s_1}$. 

Before the rejection sampling (Alg.~\ref{alg:bliss-signing}, line 5 and 6), for each signature $\bm{z}$ is distributed as:
$$
\Pr_{Z\sim D_{\sigma,( -1)^b\bm{sc}}^n} \left[Z = \bm{z} \right] =\frac{1}{\rho_\sigma(\mathbb{Z}^n)}.\exp\left(- \frac{\|\bm{z}+(-1)^b\bm{sc}\|^2}{2 \sigma^2}\right)
$$

with  $\rho_\sigma (x)$  the Gaussian function centered at 0. 
Given that the probability of rejection (Alg.\ref{alg:bliss-signing}, line 7) is :
\begin{equation*}
    \frac{1}{M. \exp\left(\frac{-\|\bm{sc}\|^2}{2 \sigma^2}\right)cosh(<\bm{z}, \bm{sc}>/\sigma^2)}
\end{equation*}
With $M$ a constant, the probability to get a specific output (b, $\bm{z}$, $\bm{c}$) is:
\begin{equation}
\label{prbability_specific_output}
    \frac{2 \exp \left(\frac{-\|z\|^2 }{2 \sigma^2}\right)}{\rho_{\sigma}(\mathbb{Z}^n)}. \frac{\exp\left(\frac{(-1)^b \langle \bm{z,\bm{sc}}\rangle}{ \sigma^2}\right)}{\exp\left(\frac{\langle \bm{z},\bm{sc}\rangle}{ \sigma^2}\right)+ \exp\left(\frac{-\langle \bm{z} \bm{sc}\rangle}{ \sigma^2}\right)}
\end{equation}
After simplifying Eqn.~\ref{prbability_specific_output}, we define in the following the likelihood function associated to one signature (b, $\bm{z}$, $\bm{c}$): 
\begin{equation}
\label{likelehood_for_one_signature}
    \hat{l}_{(b,\bm{z},\bm{c})}(\bm{s})= K+ \varphi(\langle \bm{z}, \bm{sc}\rangle)
\end{equation}
Where $K$ is constant with respect to the secret key $\bm{s}$ and $\varphi$ as follows:
\begin{equation*}
    \varphi(t)= - log\left(1+exp\left(\frac{-2t}{\sigma^2}\right)\right) 
\end{equation*} 
We are now facing the task of finding $\bm{s}$ such that we can model $N$ signatures (b, $\bm{z_k}$, $\bm{c_k}$) as accurately as possible.
Based on the likelihood function $\hat{l}_{(b, \bm{z}, \bm{c})}$ defined in Eqn.~\ref{likelehood_for_one_signature} for one signature, we calculate the log-likelihood function for $N$ given signatures, with $k<N$.
\begin{equation*}
    \hat{l}_{(\bm{b_k},z_k,c_k)}(\bm{s}) = K'+\sum_{k=0}^{N}{\varphi(\langle (-1)^b\bm{z_kc_k}, \bm{s}\rangle)}
\end{equation*}
Where $K'$ is a constant independent of $\bm{s}$.
\newline
Given that for any $\bm{w} \in  \mathbb{R}^n$ the first derivative of  $\varphi(\langle \bm{w}, \bm{s} \rangle)$ is a decreasing function and the second derivative $\varphi''$ is negative, the function is concave down and admits a maximum. Hence, $\hat{l}_{(b,z_kc_k)}$ admits a unique maximum at a point $\bm{s}$. To find out the maximum, we use gradient ascent (or descent).  

\fi

Using the trained classifier of Sec.~\ref{sec:leakage-sign-flip} and the data collected form the device under attack (Device B), we pass the public information of the signing algorithm along with the predicted values of $b_k$ (Alg.~\ref{alg:bliss-signing}, line 5) to an adaptation of \href{https://github.com/awallet/OneBitBliss}{the original code} to run the attack.

The experimental number of needed signatures is approx. 250,000 signatures for full key recovery. Alternatively, approx. 150,000 signatures are also sufficient to recover 504 coefficients of the secret key. This partial key recovery can be combined with a brute force attack to recover the remaining coefficients. 

\section{Discussion and Possible Countermeasures}
\label{sec:Discussion_and_possible_countermeasures}
In this paper, we present three power side-channel key recovery attacks on the latest constant-time implementation of BLISS ~\cite{bartheGALACTICSGaussianSampling2019}. The attacks target four subroutines: the CDT sampling, the Bernoulli rejection, the choice of sign during the Gaussian sampling, and the sign flip during signature generation. We provide sophisticated leakage analysis based on machine learning techniques. Our analysis results in revealing sensitive data that can be exploited to recover the secret key. As the targeted subroutines are also used by
other lattice-based schemes, e.g., FALCON~\cite{Fouque2019FalconFL} and FrodoKEM~\cite{NIST-FrodoKEM-20200930}, they may be vulnerable to similar attacks.

The leakage of the Bernoulli rejection alone suffices to recover the secret key by utilizing approx. 2000 signatures (Tab.~\ref{tab:attack-comparison}). Adding the use of leakages on the CDT samples $x$ and the two sign flip operations, the required number of signatures can be reduced to approx. 320. We note that the leakage of $x$ alone does not jeopardize the security of GALACTICS, as $x$ is blended with a uniformly random value. 
Therefore, we focus on discussing countermeasures against the attacks on the Bernoulli rejection and sign flip implementations.

As shown in this work, implementing Bernoulli rejection and the sign flip~\cite{bartheGALACTICSGaussianSampling2019} in constant time is not sufficient as a countermeasure against our attacks.
Masking the Bernoulli rejection and sign bit flipping might be a possible way to avoid leakage, however, d-probing models~\cite{EProuffMaskingtechnique} exploited in side-channel attacks~\cite{c31, Krachenfels2020RealWorldSV} have to be regarded. 
In~\cite{bartheGALACTICSGaussianSampling2019}, Barthe et al. proposed a  sophisticated masking technique that is claimed to be secure against higher-order side-channel attacks.  
A complete side-channel evaluation of the proposed masked implementation 
is beyond the scope of this paper. 
Yet, we believe that the masking techniques described in~\cite{bartheGALACTICSGaussianSampling2019} can be a powerful countermeasure against our attacks, namely the masked Gaussian sampler function \emph{GaussGen} and the masked random bit generator function \emph{BitGen}.

We propose partial masking techniques as a  countermeasure to the first two attacks (Alg.~\ref{attack1} and~\ref{attack2}). 
For $\yu$, the uniform value added to a random sample from the cumulative distribution table in the process of Gaussian sampling, we can predict if $\yu$ is zero or not. An effective countermeasure would be to sample $\yu$ in two halves, $y_{u1}$ and $y_{u2}$, and add them separately to the sample $Kx$.  In this case, predicting $y_{u1} = 0$ would be useless, as would $y_{u2} = 0$. Our attacks can only use leakage information when $y_{u1} + y_{u2} = 0 \mod 256$, but could only detect this if $y_{u1} = y_{u2} = 0$. This condition is unlikely to be fulfilled, which results in an increasing number of needed signatures, scaling up exponentially with the number of shares. A similar approach can be taken to mask the sign flip as proposed in ~\cite{bartheGALACTICSGaussianSampling2019}.
Here, $n$ shares are generated uniformly at random to build a Boolean sharing of a value in \{0,1\}. 

Further research is needed to investigate if and how masking can serve as an effective countermeasure against all exploitable leakage in  machine-learning-based side-channel attacks.

\section*{Acknowledgment}
The work described in this paper has been supported
 by the German Federal Ministry of Education and Research (BMBF) under the project Full Lifecycle Post-Quantum PKI - FLOQI  (ID 16KIS1074).

\bibliographystyle{alpha} 
\bibliography{biblio}



\end{document}